\documentclass{article}                            
\usepackage{amsmath,amssymb,citesort,enumerate,eucal,supertabular}

\def\bn{\bigskip\noindent}

\def\pn{\par\noindent}

\title{Giant Nemesis candidate HD 107914 / HIP 60503 for the perforation of 
Oort cloud}
\author{Igor Yu. Potemine\footnote{Universit\'e Paul Sabatier, Institut de 
Math\'ematiques de Toulouse (IMT), 118 route de Narbonne, F-31062 
Toulouse Cedex 9, France, igor.potemine@math.univ-toulouse.fr}}
\date{}

\begin{document}
\maketitle

\abstract{So far, GJ 710 is the only known star supposed to pass through 
outskirts of the solar system within 1 ly. We have reexamined the SIMBAD 
database for additional stellar candidates (from highest ratios of 
squared parallax to total proper motion) and compared them with new HIP2 
parallaxes and known radial velocities. At the moment, the best nominee 
is double star HD 107914 in the constellation Centaurus at $\approx 78.3$ pc 
from the Sun whose principal component is a white (A-type) giant. It does not 
seem to appear neither in general catalogues of radial velocities available 
at SIMBAD nor in authoritative Garc\'{\i}a-S\'anchez  et al. papers on 
stellar encounters with the solar system. Awaiting for the value $v_r$ of 
its radial velocity, unknown to the author, we have calculated limits
of $|v_r|$ necessary to this star to pass within 1 ly and 1 pc from the Sun
in linear approximation. A very accurate value of its total proper 
motion is also extremely important. In the case of $v_r=-100$ km/s and 
most ``advantageous'' HIP2 data, HD 107914 could pass as near as 8380 AU 
from the Sun in an almost direct collision course with the inner part of 
the solar system! Inversely, if $v_r$ had a great positive value, then
HIP 60503 could be the creator of peculiar trajectories of detached 
trans-Neptunian objects like Sedna.}

\section{Introduction}

Many authors (\emph{cf.}~\cite{B,D,Ga1,Ga2}) have searched past and future 
stellar perturbers of the Oort cloud. Indeed, it turns out that a 
massive star passing within 1 ly would have a significant influence on 
long-period comets. More closely, at 10000 AU, such a star would have 
a serious direct influence on trans-Neptunian objects. 

\smallskip Recently, Bobylev \cite{B} has updated the list of stars supposed 
to transit within 2 pc from the Sun using new HIP2 parallaxes. HD 107914 is 
beyond the current scope of 30 pc in Bobylev's studies. This star has two 
currently known components CCDM J12242-3855AB of visual magnitudes 7.0 and 
12.8 resp. CCDM attributes the spectral type A5 to the A-component while 
SIMBAD considers the double star as an A7/8 giant. [For the sake of 
comparison, A5-giant NSV 8327 at $\approx$ 94.9 pc has the A-component of 
spectral type A2 with $V_{\textrm{mag}}=6.0$ and B-component with 
$V_{\textrm{mag}}=14.5$ according to CCDM.] Two components are separated by 
4.6 arcsec which gives, at least, 350 AU at their current distance. The 
total mass of the system should be greater than $2M_{\odot}$.

\section{Estimations from proper motions and parallaxes}
\label{sect:estim}

We use Julian years and light-years so that 1 parsec $=p\approx 
3.261563777$ ly and 1~km/s corresponds to the velocity of 
$c^{-1}$ ly/y where $c\approx 299792.458$ is the speed of 
light in km/s.

\smallskip Let $X$ be a star with parallax $\pi_X^{}$ (in mas) at the 
current distance of $d_X=p\cdot 10^3/\pi_X$ ly from the Sun denoted
by the symbol $\odot$. The minimal distance $d_{\odot X}^{\mathrm{min}}$ 
from $X$ to $\odot$ is equal to the radius $R$ of the sphere centered at 
the Sun and tangent to the trajectory of $X$. It is easy to show that
\begin{equation}\label{eq:vrvt}
\left|\frac{v_r}{v_t}\right|=\sqrt{\left(\frac{d_X}{R}\right)^2-1}
\end{equation}
where $v_r$ and $v_t$ are radial and transverse velocities of a star $X$ 
respectively. The transverse velocity can be easily calculated from 
available catalogue values :
\begin{equation}\label{eqvt}
v_t=\frac{\pi c}{648\cdot 10^6}\mu_T^{}d_X\ \textrm{km/s}
\end{equation}
where $\mu_T^{}$ is the total proper motion of $X$ in mas and
$\frac{\pi}{648\cdot 10^6}$ the number of radians in 1 mas. From these 
two formulas we obtain
\begin{equation}\label{eq:dxmin}
d_{\odot X}^{\mathrm{min}}=\frac{d_X}{\sqrt{K^2+1}}\ \textrm{where}\ 
K=\frac{v_r}{v_t}=\frac{648\cdot 10^6}{\pi c}\times\frac{v_r}{\mu_T^{}d_X}.
\end{equation}
Supposing that $|v_r|=100$ km/s and $R\leqslant 1$ we should get
\begin{equation}
\frac{\pi_X^2}{\mu_T^{}}\gtrapprox\frac{\pi cp^2}{64800}\approx 154.6
\end{equation}
So, basically, the minimal distance to nearby stars is governed by the
$\pi_X^2/\mu_T^{}$ ratio and $\pi_X^2/\mu_T^{}\gtrsim 154.6$ is necessary 
to see a star within 1 ly from the Sun.

\smallskip For example, GJ 710 has $\pi_X^{}=51.12$, 
$\mu_\alpha^{}\cos(\delta)=1.15$, $\mu_\delta^{}=1.99$ and $v_r^{}=-13.8$
according to HIP2 and PCRV \cite{Go,vL}. So, $\pi_X^2/\mu_T^{}\approx 1137$
and $d_{\odot X}^{\mathrm{min}}\approx 0.985$ ly in a good agreement with
known predictions \cite{Ga2,B}.

\section{Calculation of limit radial velocities}\label{sect:limit}

Limit radial velocities $|v_{r,1}^{\textrm{lim}}|$ and 
$|v_{r,p}^{\textrm{lim}}|$ for stellar transits within 1 ly and 1 pc 
can be obtained from formula (\ref{eq:vrvt}) with $R=1$ and $R=p$ resp. 
HIP2 attributes $\pi_X^{}=12.77\pm 0.46$ and proper motions 
$\mu_\alpha^{}\cos(\delta)=0.55\pm 0.4$, $\mu_\delta^{}=-0.02\pm 0.3$ 
to our Nemesis candidate. [Hipparcos and SIMBAD give the following values :
$\pi_X^{}=12.89\pm 0.80$, $\mu_\alpha^{}\cos(\delta)=-0.24\pm 0.66$  and 
$\mu_\delta^{}=0.77\pm 0.52\,$.] Note that proper motions seem to swing around 
zero in different prediction modes. We summarize results in table 1 
taking into account HIP2 values and measurement errors.

\pn\tablecaption{Worst, \textbf{main} and best predictions for limit radial 
velocities necessary for HIP 60503 to pass within 1 ly and 1 pc using 
HIP2 data.}

\tablehead{\hline}\tabletail{\hline}
\begin{supertabular}{|c|c|c|c|c|c|c|}
$\pi_X^{}$  &$d_X$ &$\mu_T^{}$ &$\pi_X^2/\mu_T^{}$ &$v_t$ 
&$|v_{r,1}^{\textrm{lim}}|$ &$|v_{r,p}^{\textrm{lim}}|$\\
(mas) &(ly) &(mas/yr) &(mas$\times$yr) &(km/s) &(km/s) &(km/s)\\ 

\hline 12.31 &264.95 &1.0024 &151.17 &0.3860 &102.28 &31.36\\
\hline \textbf{12.77} &255.41 &\textbf{0.5504} &296.30 &0.2043 
&\textbf{52.18} &\textbf{16.00}\\
\hline 13.23 &246.53 &0.15 &1166.9 &0.0537 &13.25 &4.06\\
\end{supertabular}

\bn Supposing now $|v_r|=100$ km/s, we would get $d_{\odot X}^{\mathrm{min}}
\approx 0.1325$ ly $\approx$ 8380 AU in the case of ``best'' predictions! 

\section{Discussion}

Nearby stars with very small proper motions are the best targets in the
search of potential Nemeses. Our example shows that very accurate 
measurements of proper motions are indispensable for such stars.
Then one can easily create a more or less full list of Nemesis
candidates only from parallaxes and proper motions and calculate 
limit radial velocities. Finally, eliminating stars with small
radial velocities, one can use elaborated models of the galactic 
potential (\emph{cf.}~{\cite{Ga2}}) to calculate stellar trajectories and
minimal distances from target stars to the Sun.

\nocite{*}
\bibliographystyle{plain}
\bibliography{giantnem}

\end{document}